\documentstyle[twoside,fleqn,espcrc2,epsf]{article}

\newcommand{\AmS}{{\protect\the\textfont2
  A\kern-.1667em\lower.5ex\hbox{M}\kern-.125emS}}

\title{Fermion spectrum in the quenched U(1) chiral Wilson-Yukawa model}

\author{S. Aoki,\address{Institute of Physics,
        University of Tsukuba \\
        Tsukuba, Ibaraki-305, Japan}%
	\thanks{presented by S. Aoki.}
	H. Hirose$^{\rm a}$ and Y. Kikukawa$^{\rm a}$
}

\begin{document}

\begin{abstract}
Nature of charged fermion state is investigated
in the quenched U(1) chiral Wilson-Yukawa model.
Fitting the charged fermion propagator with a single hyperbolic cosine
does not yield reliable result.
On the other hand the behaviour of the propagator
including large lattice size dependence is well described by
the large Wilson-Yukawa coupling expansion,
providing strong evidence that no charged state exists as an asymptotic
particle in this model.
\end{abstract}

\maketitle

\section{INTRODUCTION}

During the past few years, the Wilson-Yukawa formulation for the
chiral gauge theories has been extensively investigated\cite{WY}.
In particular it has been found that
the fermion doublers are decoupled in the continuum limit
in the strong Wilson-Yukawa coupling region.
Recently it has been reported, however, that
no charged fermion state exists in the symmetric phase of the model
if the Wilson-Yukawa coupling is strong enough to decouple the
fermion doublers\cite{BDS}.
The conclusion is based on the numerical result obtained on a $12^3\times 16$
lattice that
the charged fermion mass is approximately equal to the sum of the neutral
fermion mass and the scalar boson mass.
We note, however, that a large finite size effect was observed for charged
fermion propagators between $8^3\times 16$ and $12^3\times 16$ lattice and
the relation among the masses does not hold on a $8^3\times 16$ lattice.
We therefore think
that much further analysis of the model,
especially a systematic study of finite
size effect, is needed. We have carried out such an analysis and we report
on the results in this talk.

\section{MODEL AND SIMULATION}

\subsection{Chiral Yukawa model}

We consider a U(1) chiral Yukawa model in the Wilson-Yukawa formulation
with the action given by $S=S_B + S_Y+S_{WY}$ where
\begin{equation}
S_B = \beta_h \sum_{n,\mu} {\rm Re}(g_n^\dagger g_n) ,
\end{equation}
\begin{eqnarray}
 S_Y   =  {1\over 2} \sum_{n,\mu}\bar\psi_n\gamma_\mu
                       (\psi_{n+\mu}-\psi_{n-\mu})
\nonumber \\
\qquad    +  Y \sum_n \bar\psi_n (g_n P_L + g_n^\dagger P_R)\psi_n
\end{eqnarray}
\begin{eqnarray}
S_{WY} = -  {r\over 2}\sum_{n,\mu }\bar\psi_n
[ P_L( (g\psi)_{n+\mu}+(g\psi)_{n-\mu}
\nonumber \\
\quad - 2(g\psi)_n ) + P_R g_n^\dagger (\psi_{n+\mu}+\psi_{n-\mu}-2\psi_n ) ] .
\end{eqnarray}
Here $g_n$ is a non-linear U(1) scalar field satisfying
$g_n^\dagger g_n =1$, $Y$  the Yukawa coupling constant and $r$ the
Wilson-Yukawa coupling. The phase transition between the broken phase and
the symmetric phase occurs at $\beta_h =0.149$ for the quenched action $S_B$.
The action $S$ is invariant under a global U(1)$_L \times$U(1)$_R$
group with U(1)$_L$ expected to be gauged in the U(1)$_L$ chiral gauge theory.
The scalar field $g_n$  has the global charged $(Q_L,Q_R)=(-1,1)$.
We define a neutral fermion operator
$N_n = ( g_n P_L + P_R)\psi_n$ having the charge $(0,1)$,
and a charged fermion operator
$C_n = ( P_L + g_n^\dagger P_R)\psi_n = g_n^\dagger N_n$ with the charge
$(1,0)$.  In terms of the neutral fermion field $N$,
the Wilson-Yukawa term  $S_{WY}$ becomes the free Wilson mass term.
Notations such as $S_F^{L}(t) =P_L S_F(t) P_R$, $S_F^{R}(t)
=P_R S_F(t) P_L$ and  $S_F^{M}(t) =P_R S_F(t) P_R+P_L S_F(t) P_L$
are used in the text where $S_F(t)$ is
a charged fermion propagator at zero momentum.

\subsection{Simulation}

Our simulation is made in the quenched approximation on a $8^3\times 16$ and
$12^3\times 16$ lattice.
Scalar field configurations are generated by the 10 hit Metropolis
algorithm at $\beta_h = 0.16$ in the broken phase
and at $\beta_h =0.145$ in the symmetric phase.

Fermion propagators are calculated by the conjugate gradient method
with the point-source in the broken phase or the wall-source in the
symmetric phase at $t=0$.
The anti-periodic (periodic) boundary condition in the time
direction (spatial directions) is used.
We fixed the Wilson-Yukawa coupling at $r=1$ and
varied the Yukawa coupling $Y$.

\section{RESULTS}

\subsection{Broken phase}

\begin{figure}[tb]
\begin{center}
\leavevmode
\epsfxsize=7.0cm
\epsfbox{fig1.ps}
\end{center}
\caption{Fitted values of $m_n$ and $m_c$ as a function of $Y$
at $\beta_h=0.16$ on a $8^3\times 16$ lattice.}
\end{figure}

In the broken phase 800 configurations separated by
200 sweeps on a $8^3 \times 16$ lattice are used to calculate
both the charged fermion propagator and the neutral fermion propagator.
The free fermion ansatz at zero spatial momentum is used to fit the
propagators in $t$ space.
The neutral fermion mass $m_n$ and the charged fermion mass $m_c$ are
plotted in fig.1 as a function of $Y$. We find that masses extracted
from the two types of propagators perfectly agree and that they also agree
with the analytic prediction of the hopping parameter
expansion(HPE)\cite{ALS} for $m_n$ .
This suggests that only the neutral fermion state exists in the broken phase.

\subsection{Symmetric phase: boson}

Our simulation in the symmetric phase is performed at $\beta_h =0.145$.
The boson mass $m_s$ is extracted through a fit of the form
$G_B(t)$=$A\cdot \cosh [m_s\cdot (8 - t)]$
where $G_B(t)$ is the scalar propagator
at zero momentum.
We obtain $m_s = 0.415(1)$ on a $8^3 \times 16$ lattice
using 1,200,000 sweeps. We have checked
that the value is insensitive to the fitting range.
The finite size effect is found to be small:  $m_s = 0.411(2)$ on a
$12^3 \times 16$ lattice. Our value of $m_s$ obtained with the
Metropolis algorithm is substantially larger than the value $m_s = 0.390(9)$
reported in ref.\cite{BDS} who used the cluster algorithm\cite{clst}.
In order to make a direct comparison
we generated 120,000 configurations with the same cluster algorithm
and found $m_s = 0.413(3)$ on a $8^3\times 16$ lattice, which perfectly
agrees with our previous value $m_s=0.415(1)$.

\begin{figure}[tb]
\begin{center}
\leavevmode
\epsfxsize=7.0cm
\epsfbox{fig2.ps}
\end{center}
\caption{Fitted value of $m_n$ as a function of $Y$ at $\beta_h=0.145$
on a $8^3\times 16$ lattice(circle) and $8^3\times 16$ lattice(square).}
\end{figure}

\subsection{Symmetric phase: fermion}

At $\beta_h = 0.145$ numbers of configurations to obtain neutral fermion
propagators are 2,400 at $Y=0.4$ and 600 at $Y=0.1$ on a
$8^3\times 16$ lattice, and 800 at $Y=0.4$ on a $12^3\times 16$ lattice.
By the free fermion ansatz the neutral fermion mass can be extracted
and
the result is given in fig 2, together with the prediction by the HPE.
It is seen that the HPE agrees with the data and
finite size effects for $m_n$ are very small.

\begin{figure}[bt]
\begin{center}
\leavevmode
\epsfxsize=7.5cm
\epsfbox{fig3.ps}
\end{center}
\caption{Fitted value of $m_c$ as a function of $t_{min}$
at $\beta_h=0.145$ and $Y=0.4$
on a $8^3\times 16$ lattice(circle) and $12^3\times 16$ lattice(square).}
\end{figure}

At $\beta_h=0.145$ charged fermion propagators
are averaged over
2,400 (3,600) configurations separated by 500 sweeps
on a $8^3 \times 16$ ($12^3\times 16$) lattice at $ Y = 0.1$ and $0.4$.
We found that $S_F(t)$ can not be fitted by the free fermion ansatz.
We therefore tried to extract the charged fermion mass through a fit of the
form
\begin{equation}
S_F(t)= A\cdot \cosh [m_c\cdot (8-t)]  \label{eq:fit}
\end{equation}
in the range $t_{min}\leq t\leq 16-t_{min}$.
The value $m_c$ as a function of $t_{min}$ is plotted in fig.3.
Since the value of $m_c$ significantly changes with the variation of $t_{min}$,
the charged fermion mass can not be extracted reliably by the fit.
We are therefore not able to confirm the relation claimed in ref.\cite{BDS}
that $m_c \simeq m_n + m_s$ on a $12^3\times 16$ lattice.
We should comment that our $S_F(t)$ numerically differs from that of
ref.\cite{BDS} by $3\sim 5$ standard deviations,
exhibiting a faster decrease with $t$.
This discrepancy probably originates from the difference of boson
configurations noted in Sec.3.2.
We have attempted to fit the data of ref.\cite{BDS} by
eq.(\ref{eq:fit}),
finding, however, the behaviour of $m_c$ similar to that of fig.3
for a $12^3\times 16$ lattice.

\begin{figure}[tb]
\begin{center}
\leavevmode
\epsfxsize=7.5cm
\epsfbox{fig4.ps}
\end{center}
\caption{$S_F^{L}(t)$ on $8^3\times 16$ lattice(circle) and
$12^3\times 16$ lattice(square), together with the analytic predictions
at the leading order(dashed lines) and at the next leading order(solid lines),
at $\beta_h=0.145$ and $Y=0.4$.}
\end{figure}

\subsection{Finite size effect}

At $\beta_h = 0.145$ a large finite size effect is observed between two
lattice sizes.
After the conference we have analyzed this finite size effect.
We performed the large Wilson-Yukawa coupling
expansion\cite{GPR} up to the next-to-leading order
and compared the result with the numerical data.
The data and the analytic prediction for $S_F^{L}(t)$ on
two lattice sizes ($8^3\times 16$ and $12^3\times 16$) are plotted
in fig.4 ($ Y =  0.4$) and fig.5 ($ Y = 0.1$).
The agreement between the data and the prediction is excellent
at $Y=0.4$ and is very good even at $Y=0.1$.
The leading order contribution for $S_F^{L}(t)$
in this expansion is given by a convolution of the neutral fermion
propagator and the scalar propagator, each of which has very small lattice
size dependence. However it turns out that their convolution has
large size dependence, which explains the behaviour of our data, as seen
in fig.4 and fig.5.
The next-leading order correction to the leading order is relatively small.
The analytic prediction for $S_F^{R}(t)$ and $S_F^{M}(t)$ also agrees with the
data very well\cite{AHK}.

\begin{figure}[tb]
\begin{center}
\leavevmode
\epsfxsize=7.5cm
\epsfbox{fig5.ps}
\end{center}
\caption{Same as Fig.4 at $\beta_h=0.145$ and $Y=0.1$.}
\end{figure}

In the large Wilson-Yukawa coupling expansion
it has been shown\cite{GPR} that
the charged fermion is a scattering state of the scalar boson and the neutral
fermion.
The agreement between the data and the result of the expansion
provides strong evidence that no charged fermion exists as a single particle
state in the symmetric phase.

\section{CONCLUSION}
We have investigated the spectrum of fermions in the Wilson-Yukawa formulation
through a quenched simulation of the U(1) chiral Wilson-Yukawa model.
In the broken phase
the neutral fermion state exists and the charged fermion operator generates
the same state.
In the symmetric phase
the charged fermion mass can not be extracted by the single hyperbolic cosine
fit at large $t$ and charged fermion propagators suffer
a large finite size effect.  We found that not only this large finite
size effect but also the form of charge fermion propagators
can be explained by the large Wilson-Yukawa coupling expansion.
In this expansion the charged fermion is interpreted as a scattering state
of the scalar boson and the neutral fermion.
We conclude that no charged fermion exists
as a single particle asymptotic state in the Wilson-Yukawa
model in {\it both} phases.
Only the scalar boson and the neutral fermion exist in this model.

\section*{ACKNOWLEDGEMENTS}

Numerical calculations for the present work were carried out on HITAC S820/80
at KEK.
We thank the Theory Division of KEK for warm hospitality.
We also thank Prof. Kanaya and Prof. Ukawa for valuable discussions.
Y.K. is supported by a Grant-in-Aid for JSPS fellow.

\end{document}